\documentclass{IEEEtran}
\pdfoutput=1
\usepackage{graphicx}
\usepackage{amsmath}
\usepackage{latexsym}
\usepackage{amssymb}
\usepackage{subfigure}
\usepackage{slashbox}
\usepackage{hyperref}
\usepackage{cite}
\usepackage{algorithm}
\usepackage{algorithmic}
\usepackage{color}
\usepackage{setspace}

\begin{document}
%
\title{Joint Sparse Recovery with Semi-Supervised MUSIC}
\author{Zaidao Wen,~Biao Hou,~\emph{Member,~IEEE},~Licheng Jiao,~\emph{Senior Member,~IEEE}
}
\maketitle

\begin{abstract}
Discrete multiple signal classification (MUSIC) with its low computational cost and mild condition requirement becomes a significant non-iterative algorithm for joint sparse recovery (JSR). However, it fails in rank defective problem caused by coherent or limited amount of multiple measurement vectors (MMVs). In this paper, we provide a novel sight to address this problem by interpreting JSR as a binary classification problem with respect to atoms. Meanwhile, MUSIC essentially constructs a supervised classifier based on the labeled MMVs so that its performance will heavily depend on the quality and quantity of these training samples. From this viewpoint, we develop a semi-supervised MUSIC (SS-MUSIC) in the spirit of machine learning, which declares that the insufficient supervised information in the training samples can be compensated from those unlabeled atoms. Instead of constructing a classifier in a fully supervised manner, we iteratively refine a semi-supervised classifier by  exploiting the labeled MMVs and some reliable unlabeled atoms simultaneously. Through this way, the required conditions and iterations can be greatly relaxed and reduced. Numerical experimental results demonstrate that SS-MUSIC can achieve much better recovery performances than other MUSIC extended algorithms as well as some typical greedy algorithms for JSR in terms of iterations and recovery probability. The code is available on \url{https://github.com/wzdammy/semi_supervised_MUSIC}.
\end{abstract}


%
\IEEEpeerreviewmaketitle

\section{Introduction}
\label{sec:intro}
\PARstart{T}{he} emerging theory of compressed sensing (CS) supplies a paradigm for recovering an unknown sparse signal from some compressed linear measurements and it has been devoted to many applications in signal processing as well as machine learning  (ML) \cite{donoho2006compressed,baraniuk2007compressive,candes2008introduction}. This theory primarily addresses the recovery problem of a signal $\mathbf{x}\in\mathbb{R}^n$ from its single measurement vector (SMV) $\mathbf{y}\in\mathbb{R}^{m}$ such that $\mathbf{y}=\mathbf{A} \mathbf{x}$, where $\mathbf{A}\in\mathbb{R}^{m\times n}$ models the linear measurement matrix with $m\ll n$. Practically, we may encounter the problem of simultaneously recovering a group of $N$ sparse signals $\mathbf{X}=[\mathbf{x}_1,\dots,\mathbf{x}_N]\in\mathbb{R}^{n\times N}$ from their multiple measurement vectors (MMVs) $\mathbf{Y}$  in many tasks, \emph{e.g.,} multivariate regression \cite{Peng2009}, classification \cite{Wen2016,Li2015}, direction of arrival estimation \cite{Malioutov2005}, \emph{etc}. When these underlying signals share some particular sparse patterns, it will enable to reduce the condition for successful recovery. One of the most prevalent patterns expressed as joint sparse suggests that these signals will share the same support so that $\mathbf{X}$ will contain only a few non-zero rows. In this scenario, if the row-sparsity, the number of non-zero rows of $\mathbf{X}$ is equal to $K$, the problem of joint sparse recovery (JSR) from a common $\mathbf{A}$ can be formulated as
\begin{equation}\label{Equ:MMVproblem}
  \min_{\mathbf{X}} \|\mathbf{Y-AX}\|_\mathrm{F}^2,\quad\mathrm{s.t.}~\|\mathbf{X}\|_{\mathrm{row},0}\leq K,
\end{equation}
where $\|\cdot\|_\mathrm{F}$ is the Frobenius norm (F-norm) and $\|\mathbf{X}\|_{\mathrm{row},0}$ counts the non-zero rows in $\mathbf{X}$. Unfortunately, \eqref{Equ:MMVproblem} is generally a combinatorial non-convex optimization problem due to $\|\mathbf{X}\|_{\mathrm{row},0}$. To solve this problem, two strategies have been developed in optimization field, namely convex relaxation with a mixed norm and greedy methods \cite{Hyder2009}. Focusing on the greedy algorithm, the central issue becomes to iteratively estimate a certain amount of atoms according to the correlations with the residual matrix so as to mostly decrease the value of objective function \eqref{Equ:MMVproblem}. Once a support set is determined, the recovery problem will be reduced to a standard overdetermined linear problem solved with the least square.  As a consequence, numerous greedy JSR algorithms have been extended from SMV to MMVs \cite{Blanchard2014Greedy}, yielding the orthogonal matching pursuit (OMP) for MMVs (OMP-MMV) \cite{tropp2004Greed,Chen2006} or the so-called simultaneously OMP (SOMP) \cite{Tropp2006,Determe2016}, simultaneously compressive sampling matching pursuit (SCoSaMP) \cite{needell2009cosamp,Blanchard2014Greedy}, rank aware order recursive matching pursuit (RA-ORMP) \cite{Davies2012Rank} \emph{etc}.
\par Another significant algorithm referred to as discrete multiple signal classification (MUSIC) takes a different viewpoint in the field of signal processing \cite{Feng1996Spectrum}. It reveals that each measurement vector and the correct atoms should reside in the same subspace if $\mathrm{rank}(\mathbf{Y})\doteq r=K$. Under a mild condition, those atoms can be straightforward determined by singular value decomposition (SVD) without iterative process, which achieves far more remarkable performance than those greedy optimization algorithms in terms of complexity and required conditions. When $r<K$ caused by limited amount of MMVs or information loss due to their correlations, MUSIC will however yield a failing estimation in this rank defective case. {To overcome this drawback, some MUSIC extended algorithms have been developed for rank defective problem, such as iMUSIC, compressive MUSIC (CS-MUSIC) and subspace-augmented MUSIC (SA-MUSIC) \cite{Lee2010,Kim2012,Lee2012}. iMUSIC, an initial version of SA-MUSIC, involves an iterative atom refinement procedure in MUSIC so that some falsely determined atoms could be gradually refined during iterations. However, some operations in atoms refinement are not optimal so that it can only achieve a marginal improvement than MUSIC and some conventional greedy approaches in noiseless case. Later, two almost equivalent algorithms of SA-MUSIC and CS-MUSIC provide a two-stages framework, which indicates that if any $K-r$ atoms could be correctly estimated in the first stage with any an off the shelf algorithm, the rest $r$ atoms will be simply determined by applying MUSIC on an augmented subspace \cite{Kim2012,Lee2012}. It follows that the required conditions and iterations for such a combinational framework will actually depend on the algorithm in the first stage, which is usually suboptimal compared to MUSIC. How to fully exploit the advantages of MUSIC to relax the condition and reduce the iterations become two important issues.}
\par {In recent years, the field of ML attracts much more attentions because of some significant progresses in both theory and industry. If we revisit the support estimation from the perspective of ML, it can be regarded as a binary classification task with respect to atoms and MUSIC actually constructs a nearest subspace classifier (NSC) according to the positive labeled training samples $\mathbf{Y}$ in a fully supervised way \cite{Liu2011}. Therefore, its discriminative ability will be naturally affected by the quality and quantity of these training samples. Following this novel viewpoint, we are motivated to address the rank defective problem by means of the strategy in ML. }
\par {In this paper, we present a novel semi-supervised MUSIC (SS-MUSIC) for JSR, in which both the labeled MMVs and some reliable unlabeled atoms are iteratively exploited for classifier construction \cite{Topor2009Semi}. Through this way, the inadequate supervised information in rank defective MMVs can be additionally compensated from those unlabeled data so as to increase the discrimination of the classifier. As a consequence, SS-MUSIC will successfully classify all atoms within $K-r$ iterations as long as at least one positive atom can be newly determined and preserved in each iteration. The simulation results clearly demonstrate the superiorities of SS-MUSIC, compared with the other MUSIC extended frameworks as well as some state-of-the-art greedy algorithms.}
\par The rest paper is organised as follows. Sec. \ref{sec:Proposed Method} proposes our algorithm in detail. Numerical experiments are conducted in Sec. \ref{sec:experiment} and Sec. \ref{sec:Conclusion} concludes this paper.

\section{Semi-Supervised MUSIC}\label{sec:Proposed Method}
\par In this section, we will formally reformulate the JSR problem and MUSIC from the viewpoint of ML in the first place. Then a novel SS-MUSIC framework is developed and compared with the other algorithms in order to demonstrate its superiorities.

\subsection{{Reformulation of JSR and MUSIC}}
\par Let $\mathbf{Y}\in\mathbb{R}^{m\times N}$ contain $N$ labeled noiseless training samples drawn from the positive class. Given $n$ unlabeled atoms $\{\mathbf{a}_i\}_{i=1}^n$, the central task for JSR is classifying these atoms into two classes by assigning a proper label $l_i\in\{0,1\}$ to $\mathbf{a}_i$ such that $\mathbf{Y}=\sum_{i}\delta(l_i)\mathbf{a}_i\mathbf{X}^i$, where $\mathbf{X}^i$ is the $i$-th row vector in $\mathbf{X}$, $\delta$ stands for the indicator function as $\delta(l_i=0)=0$ for negative label and $\delta(l_i=1)=1$ for positive one. Additionally, we have a prior knowledge that the amount of the positive atoms will be $K$. {Since each $\mathbf{y}_i$ will reside in the subspace $\mathcal{S}_{\mathbf{A}(\mathbf{l}_+)}$ spanned by those positive labeled atoms, we can measure the sum of Euclidean distance from each $\mathbf{y}_i$ to $\mathcal{S}_{\mathbf{A}(\mathbf{l}_+)}$ to evaluate the fitness of a label configuration $\mathbf{l}\in\{0,1\}^n$, which is defined as following.}

{\begin{equation}\label{Loss}
\setlength{\abovedisplayskip}{1pt}
  \ell(\mathbf{l})=\sum_{i=1}^N \mathrm{dis}(\mathbf{y}_i,\mathcal{S}_{\mathbf{A}(\mathbf{l}_+)})=\sum_{i=1}^N\|\mathbf{y}_i-\mathcal{P}_{\mathcal{S}_{\mathbf{A}(\mathbf{l}_+)}}(\mathbf{y}_i)\|_2^2
\end{equation}
where $\mathbf{A}(\mathbf{l}_+)$ stands for the subset of atoms with positive labels and $\mathcal{P}_{\mathcal{S}}$ is the orthogonal projection operator onto subspace $\mathcal{S}$ here and after. If $\mathbf{U}$ is the orthogonal basis of subspace $\mathcal{S}$ computed from truncated SVD or principal component analysis (PCA) \cite{smith2002tutorial}, $\mathcal{P}_{\mathcal{S}}=\mathbf{UU}^\mathrm{T}$. It follows that if the classification is correct, $\ell(\mathbf{l})$ will reach its lower bound, namely zero in a noiseless situation. In practical situation with noisy MMVs, we can exploit a threshold $\epsilon$ related to signal-to-noise ratio (SNR) to indicate the fitness of $\mathbf{l}$, namely $\ell(\mathbf{l})\leq \epsilon$.}
\par {Considering this task, one of the most prevalent strategies in ML is supervised classification, which focuses on constructing a classifier based on the training samples. Following this way, the novel MUSIC algorithm essentially constructs a NSC with $\mathbf{Y}$ and each query atom will be classified by measuring $\mathrm{dis}(\mathbf{a}_i,\mathcal{S}_\mathbf{Y})$ \cite{Feng1996Spectrum}. Then $K$ atoms with the closest distance will be classified into the positive class. Since this classifier will be frequently exploited in the following paper, it will be denoted by $\mathbf{l}\gets\mathcal{W}(\mathcal{Q}|\mathcal{S}_{\mathcal{C}},K)$, where $\mathcal{Q}$ contains the query testing samples, $\mathcal{S}_{\mathcal{C}}$ is the subspace spanned by samples in set $\mathcal{C}$ and $K$ controls the number of positive labels in the output of label configuration $\mathbf{l}$. }
\begin{figure}
  \centering
  \subfigure[]{\includegraphics[width=0.18\textwidth]{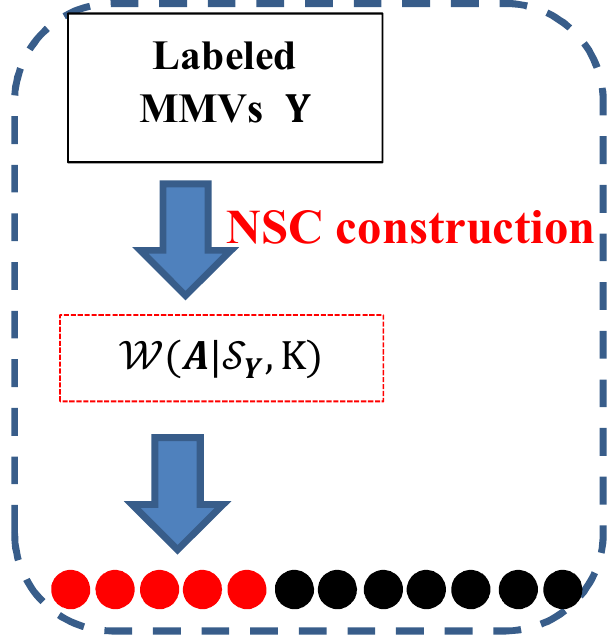}\label{Fig:MUSIC}}
  \subfigure[]{\includegraphics[width=0.29\textwidth]{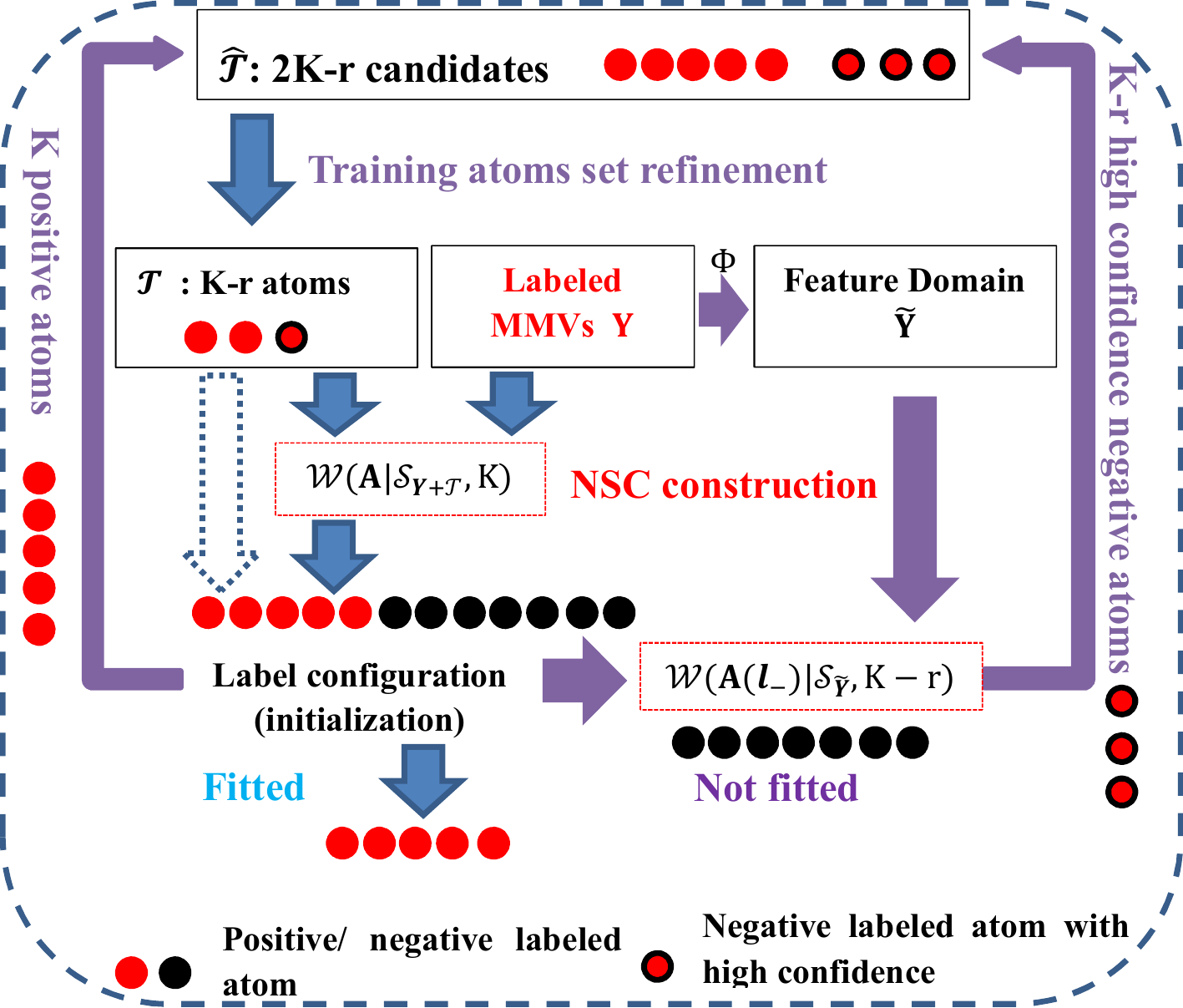}\label{Fig:SSMUSIC}}
  \caption{Framework illustrations. (a). MUSIC and (b). SS-MUSIC.}\label{Fig:MUSICillustration}
\end{figure}
\subsection{Algorithm Presentation}
\par It has been indicated that when $\mathrm{rank}(\mathbf{Y})=K$, such a supervised classifier will produce a perfect classification result if any $K+1$ atoms are linearly independent \cite{Feng1996Spectrum}. However, when the number of training samples is limited or they are coherent, $\mathrm{rank}(\mathbf{Y})<K$. In this case, supervised information in training data will be insufficient to construct a discriminative NSC so that the performance will be degraded. To overcome this deficiency, we will consider the strategy of semi-supervised classification (SSC) to develop a novel SS-MUSIC framework \cite{Topor2009Semi}, whose central idea is to simultaneously make use of the labeled and some reliable unlabeled samples to construct a semi-supervised classifier. Then the information required for classification will be compensated from unlabeled data. For better understanding the difference between MUSIC and SS-MUSIC, two frameworks will be illustrated in Figs.\ref{Fig:MUSICillustration}, where the notation $\mathcal{\widehat{T}}$ and $\mathcal{T}$ containing unlabeled atoms will represent the candidate and actual training set for semi-supervised classifier construction, respectively. we will address the two central issues of constructing $\widehat{\mathcal{T}}$ and $\mathcal{T}$ to explain the framework in Fig. \ref{Fig:SSMUSIC} as following.

\par We will start from a label configuration $\mathbf{l}^t$ obtained in $t$-th iteration, $t\geq 0$. If it is not fitted according to \eqref{Loss}, it implies that the current training set cannot provide sufficient or correct discriminative information for classifier construction, \emph{i.e.}, $\mathcal{T}^t$ will contain some outlying atoms so as to bias the classifier or the number of involved atoms is inadequate. To address this issue, except for those atoms in current positive class, we will reappraise the confidence of each atom in the negative class so that some with high confidences will be also involved in $\widehat{\mathcal{T}}^{t+1}$. We suggest that if an atom is much similar to $\mathbf{Y}$ measured in a feature domain, a high confidence of being involved will be encouraged. To avoid the redundant information, the high confident atoms should have the ability of providing extra information compared with $\mathbf{A}_{\mathbf{l}_+^t}$. {To meet these two requirements, we will firstly project $\mathbf{Y}$ onto the orthogonal complement subspace of $\mathbf{A}(\mathbf{l}_{+}^t)$ as $\widetilde{\mathbf{Y}}=\Phi(\mathbf{Y})$ with a feature extractor $\Phi$ in order to eliminate the information of $\mathbf{A}(\mathbf{l}_+^t)$. Then NSC will be exploited to select $K-r$ atoms $\mathbf{\widehat{A}}(\mathbf{l}_{-}^t)$ that is nearest to $\mathcal{S}_{\widetilde{\mathbf{Y}}}$. Finally, the candidate set will be constructed as $\widehat{\mathcal{T}}^{t+1}\gets \{\mathbf{A}(\mathbf{l}_{+}^t),\mathbf{\widehat{A}}(\mathbf{l}_{-}^t)\}$. The above procedures are denoted by purple flows in Fig. \ref{Fig:SSMUSIC}.}

\par After we obtain $\widehat{\mathcal{T}}^{t+1}$ containing $2K-r$ candidates, $K-r$ representative and reliable atoms will be further refined to update $\mathcal{T}^{t+1}$ and construct the semi-supervised classifier. This task will be simply interpreted as the following overcomplete variables selection problem \cite{Crocker1980Linear}.
\begin{equation}\label{Equ:Selection}
  \mathbf{\widehat{X}}^*\gets\arg\min_{\widehat{\mathbf{X}}} \|\mathbf{Y}-\mathbf{A}_{\widehat{\mathcal{T}}}\mathbf{\widehat{X}}\|_\mathrm{F}^2
\end{equation}
whose solution is given by $\mathbf{\widehat{X}}^*=\mathbf{A_{\widehat{\mathcal{T}}}}^{\dag}\mathbf{Y}$ and $\mathbf{A_{\widehat{\mathcal{T}}}}^{\dag}$ stands for the pseudo-inverse of $\mathbf{A_{\widehat{\mathcal{T}}}}$. Then the atoms corresponding to the first $K-r$ largest $\|(\mathbf{\widehat{X}}^*)^j\|_2$ will be selected into $\mathcal{T}^{t+1}$. Next, atoms in $\mathcal{T}^{t+1}$ and the labeled MMVs will be simultaneously used to construct a semi-supervised classifier as  $\mathbf{l}^{t+1}\gets\mathcal{W}(\mathbf{A}|\mathcal{S}_{\mathcal{T}^{t+1}+\mathbf{Y}},K)$. Since $K-r$ atoms are already devoted to classifier construction, their labels will be consequently positive and we only need to assign the rest $r$ positive labels to other atoms in $\mathbf{A}$.  The complete SS-MUSIC is summarised in following Algorithm \ref{Algorithm1}.
\begin{algorithm}[b]
\caption{Semi-Supervised MUSIC for Noiseless JSR}\label{Algorithm1}
\begin{algorithmic}[1]
\STATE \textbf{Input}: Row Sparsity: $K$;~
Measurement Matrix $\mathbf{A}$;~MMVs $\mathbf{Y}$;~Threshold: $\epsilon$; Maximum iterations:  $T_{max}$.
\STATE{ Initialization: $\mathbf{l}^{0}=\mathbf{0}$, $\mathcal{T}^{0}=\emptyset$, $t\gets0$,~$r=\mathrm{rank}(\mathbf{Y})$.}
 \WHILE{$\ell(\mathbf{l}^t)>\epsilon$ or $t<T_{max}$}
 \STATE {Project $\mathbf{Y}$ onto a feature domain as $\mathbf{\widetilde{Y}}=\Phi(\mathbf{Y})$.}\\
  \STATE {Compute $\mathbf{\widehat{A}}(\mathbf{l}_{-}^t)$ and update $\widehat{\mathcal{T}}^{t+1}\gets \{\mathbf{A}(\mathbf{l}_{+}^t),\mathbf{\widehat{A}}(\mathbf{l}_{-}^t)\}$.}\\
 \STATE {Update $\mathcal{T}^{t+1}$ to construct the semi-supervised classifier as $\mathbf{l}^{t+1}\gets\mathcal{W}(\mathbf{A}|\mathcal{S}_{\mathcal{T}^{t+1}+\mathbf{Y}},K)$ .}\\
 \ENDWHILE
\end{algorithmic}
\end{algorithm}
\subsection{{Discussion and Comparison}}
\par To demonstrate the superiorities of SS-MUSIC to make it more convinced, some discussions and comparisons with other algorithms will be carried out in this subsection, in spite of their distinct motivations. In the first place, let us focus on the iterations. According to the principle of SA-MUSIC or CS-MUSIC, once $\mathcal{T}^{t+1}$ has contained the $K-r$ atoms belonging to the positive class, the subsequent $\mathcal{W}(\mathbf{A}|\mathcal{S}_{\mathcal{T}^{t+1}+\mathbf{Y}},K)$ will generate the fitted label configuration. It follows that if one more positive atom could be newly involved and preserved in $\mathcal{T}$ in each iteration, the upper bound on iterations will be $K-r$. In fact, we will empirically show in the next section that the actual iterations will be much fewer than $K-r$. On the contrary, CS-MUSIC and SA-MUSIC respectively exploit M-OMP and OSMP to determine $K-r$ atoms iteratively in the first stage so that their iterations will be always $K-r$. Since iMUSIC also adopts M-OMP for initial $K-r$ atoms estimation, the lower bound on iterations will be $K-r$. Accordingly, SS-MUSIC requires fewer iterations than these MUSIC extended algorithms while its computational complexity will be still comparable with that of iMUSIC. Now let us compare the required conditions for each algorithm. For SA-MUSIC and CS-MUSIC, their required conditions mainly come from the first stage that should guarantee the correctness of selecting one atom in each iteration. On the contrary, SS-MUSIC will only require at least one correct atom to be selected and preserved in $\mathcal{T}$, which will be reasonably much relaxed than that of SA-MUSIC and CS-MUSIC. This can be also concluded from the proof of the generalized OMP which selects a set of atoms in one iteration to relax the condition of OMP in SMV problem \cite{wang2012generalized}. Additionally, SS-MUSIC involving an atom refinement process will further relax the conditions, which is similar to iMUSIC. Nevertheless, SS-MUSIC is different from iMUSIC in following implementations. 1). iMUSIC utilizes $\mathcal{W}(\mathbf{A}(\mathbf{l}_{-})|\mathcal{S}_{\mathcal{T}^{t}+\mathbf{Y}},K_0)$ to construct $\mathcal{\widehat{T}}$, where $K_0$ is the number of selected atoms controlled by the condition number of $\mathbf{A}_{\widehat{\mathcal{T}}}$. On the contrary, SS-MUSIC selects $K-r$ atoms based on $\mathcal{W}(\mathbf{A}(\mathbf{l}_{-})|\mathcal{S}_{\widetilde{\mathbf{Y}}},K-r)$ in a different feature subspace. 2). Eq. \eqref{Equ:Selection} in iMUSIC is different and the resulted $K$ atoms will be directly served as the label configuration in this iteration. In SS-MUSIC, those resulted $K-r$ atoms will be subsequently utilized to build a semi-supervised classifier $\mathcal{W}(\mathbf{A}|\mathcal{S}_{\mathcal{T}^{t+1}+\mathbf{Y}},K)$ to obtain the label configuration.
\par Comparing SS-MUSIC with some conventional greedy optimization algorithms in CS, it will be analogous to SCoSaMP which estimates $2K$ atoms and makes a refinement in each iteration. Nevertheless, SCoSaMP requires a more strict condition and more iterations for exact recovery. Ambat and Hari presented a general iterative framework for SMV problem \cite{Ambat2015}, in which they introduce a regularization procedure on both atoms and measurement vector to remove the effect of the previous estimated atoms. However, they aim at estimating complete $K$ atoms in each iteration while SS-MUSIC focuses on $K-r$ in the spirit of SSC.
\section{{Empirical Performance}}
\label{sec:experiment}
In this section, we consider the following experimental setting to evaluate the performance for rank defective JSR problem. $\mathbf{X}\in\mathbb{R}^{100\times N}$ is drawn from the standard Gaussian distribution and $K>N$ rows in general position are randomly retained with other rows setting as zeros. In this case, $\mathrm{rank}(\mathbf{Y})=r=N$. $\mathbf{A}\in\mathbb{R}^{m\times 100}$ is also chosen as the standard Gaussian random matrix with each atom normalized. $T_{max}=100$.
\par
\begin{figure}
  \centering
  \subfigure[]{\includegraphics[width=0.11\textwidth]{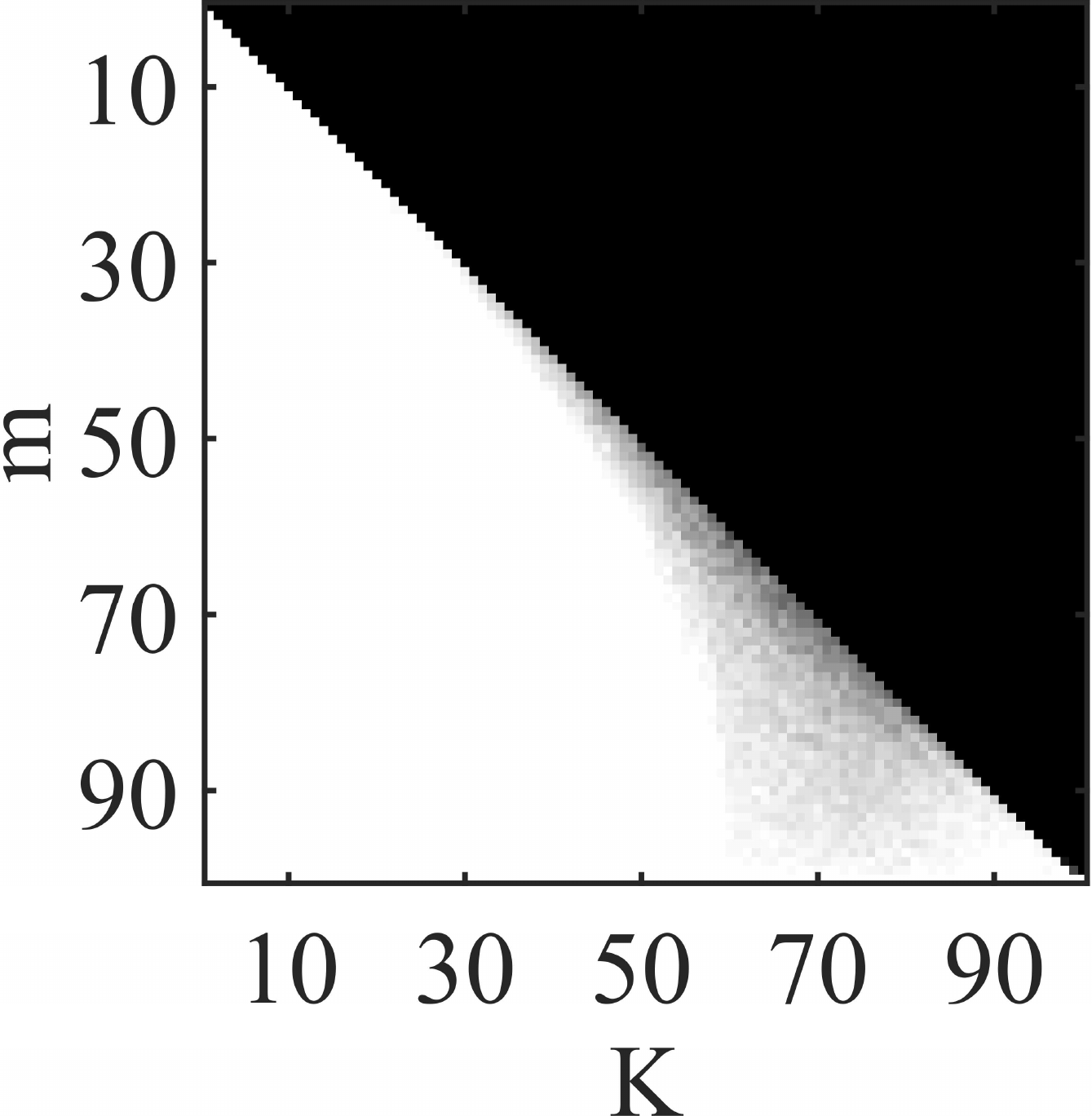}\label{Fig:phasetransition}}
  \subfigure[]{\includegraphics[width=0.15\textwidth]{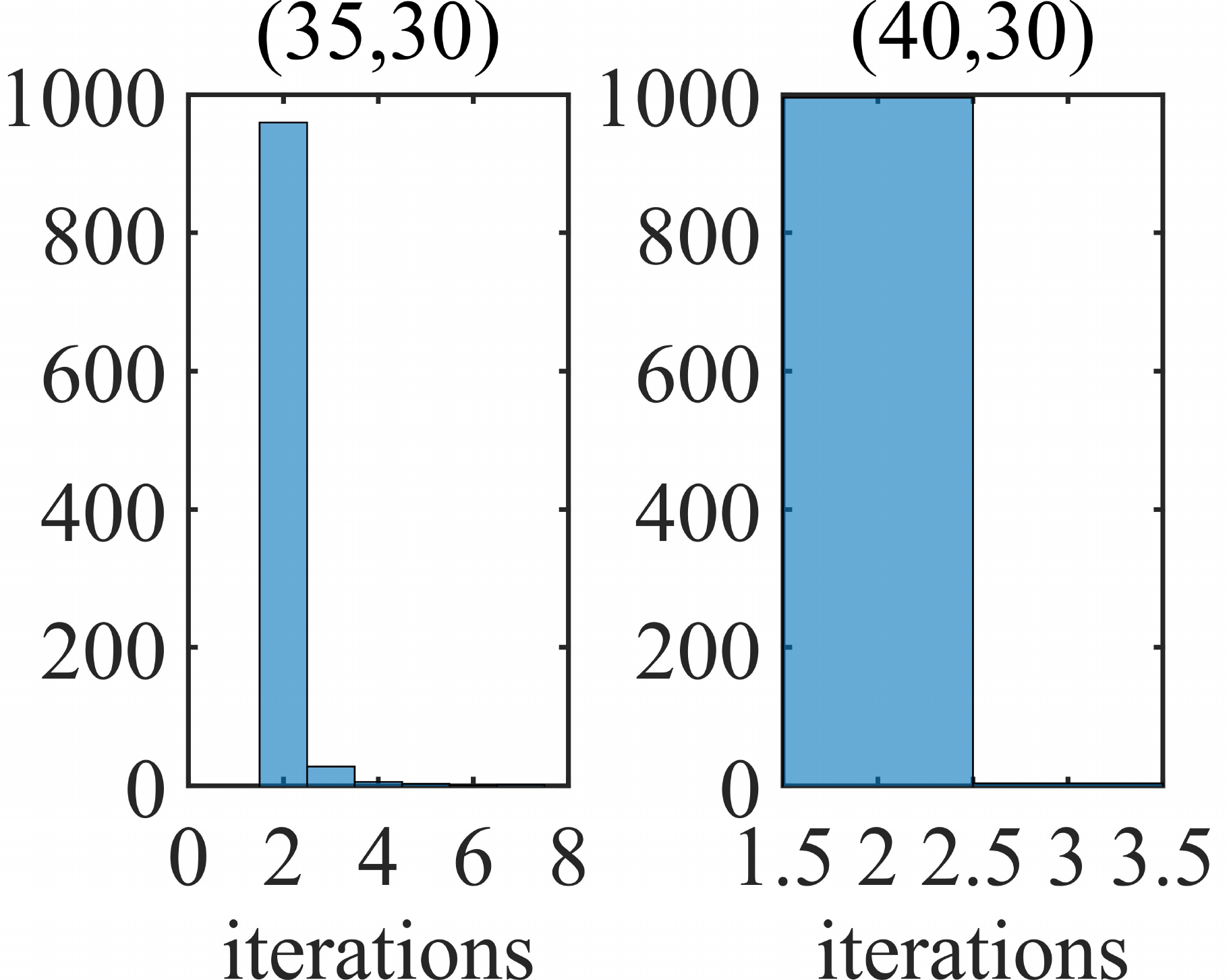}\label{Fig:ssmusichist}}
    \subfigure[]{\includegraphics[width=0.15\textwidth]{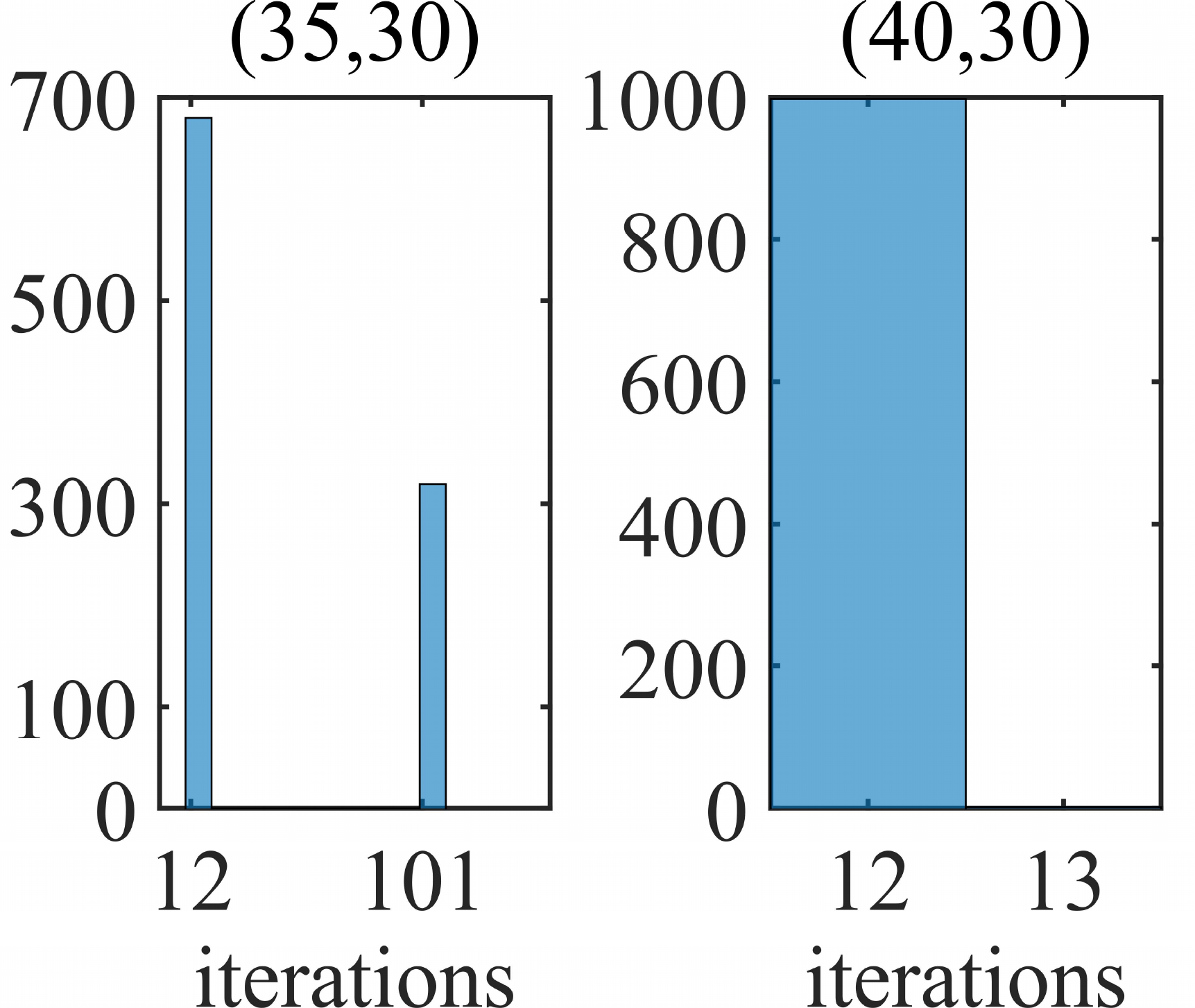}\label{Fig:imusichist}}
  \caption{(a). Phase transition maps. (b) Histograms of iterations for SS-MUSIC. (c) Histograms of iterations for iMUSIC. }\label{Fig:phasetransition}
\end{figure}

 In the first part, the phase transition of SS-MUSIC will be evaluated in Fig. \ref{Fig:phasetransition}, where $m$ and $K$ will vary from 1 to 100, respectively and $N=20$ and 1000 independent recovery experiments are conducted for each pair of $(m,K)$. We can observe from the transition map that when $m>K$, SS-MUSIC will perfectly recover the signals with high probabilities for most pairs of $(m,K)$. When $K<20$, it becomes the full rank JSR problem, in which case SS-MUSIC will be the MUSIC sharing condition of $m=K+1$. When $K>20$, we however observe that the required $m$ for recovery will not greatly increase to reach the high probability performance, which empirically demonstrates a mild condition for $\mathbf{A}$. To evaluate the required iterations, we choose the recovery results from pair $(35,30)$ and $(40,30)$ and count the iterations for successfully recovery. For those failure trials, the iterations will be denoted by $101$ as $T_{max}=100$. We plot the histograms of the iterations in Fig. \ref{Fig:ssmusichist}. For comparison, the results for iMUSIC will be also illustrated in Fig. \ref{Fig:imusichist}. It should be declared that the standard iMUSIC leverages the condition number to control the amount of involved atoms for refinement, but it will involve another parameter. Instead, a fixed number is selected which is the same with SS-MUSIC, namely $K-r$. We can see from the results in Fig. \ref{Fig:ssmusichist} that SS-MUSIC can achieve perfect performances in both cases and the required iterations can reach the lower bound 2 in the most experiments among 1000 trials. On the contrary, 300 trials for iMUSIC in the case of $(35,30)$ are failed and the iterations in the rest experiments are 12.
\par
\begin{figure}
  \centering
  \subfigure[]{\includegraphics[width=0.15\textwidth]{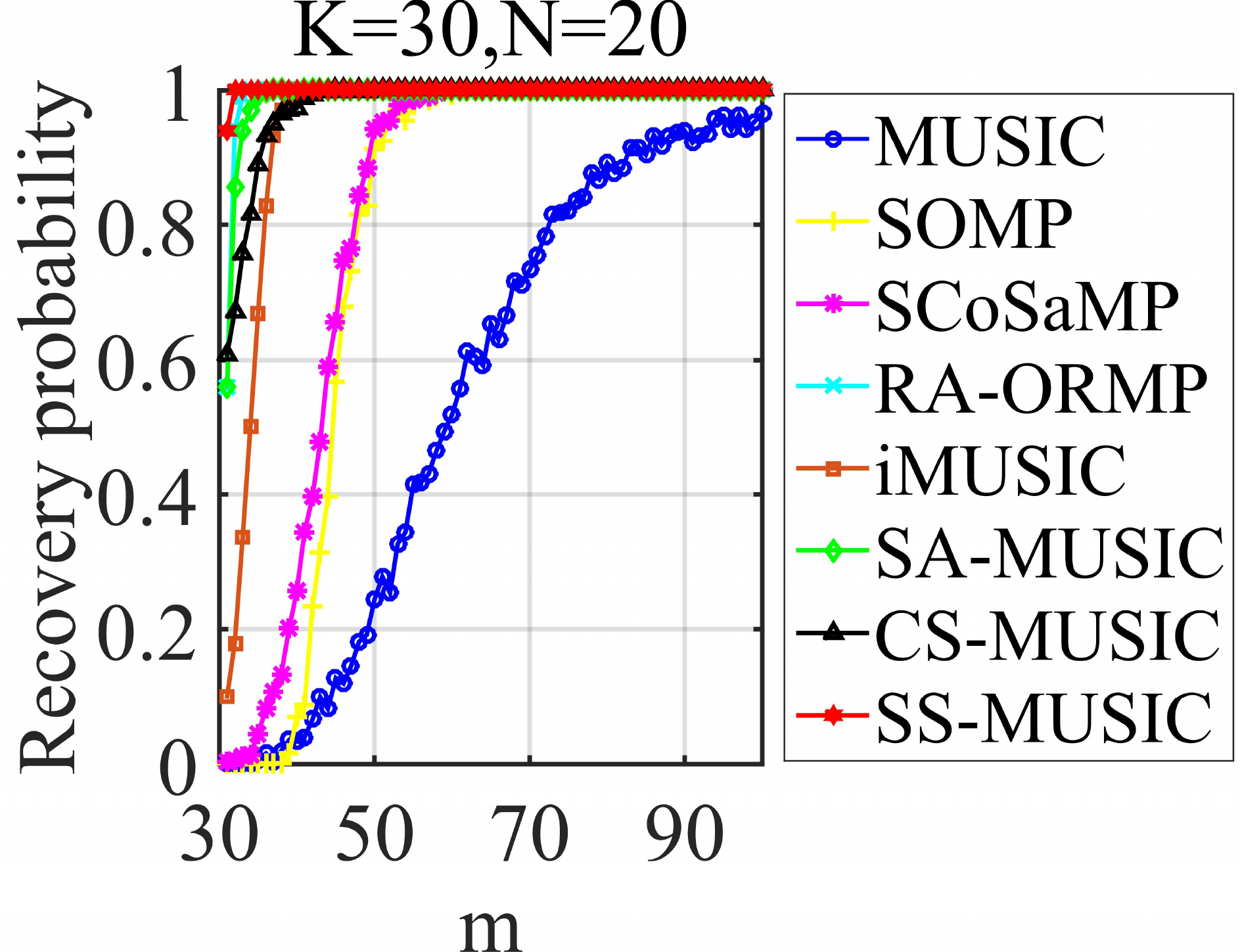}\label{RecoverProb_measurement}}
   \subfigure[]{\includegraphics[width=0.15\textwidth]{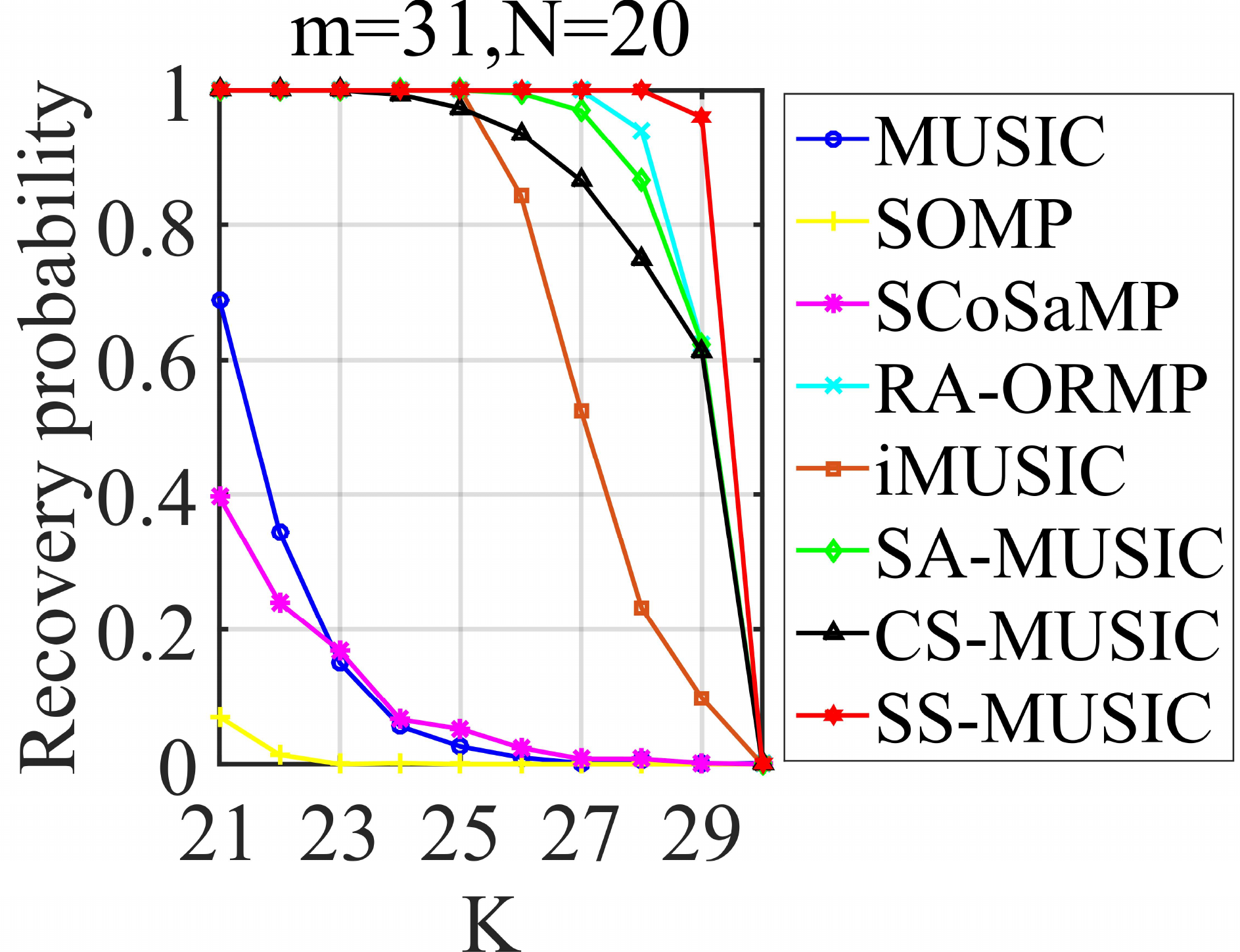}\label{RecoverProb_sparsity}}
    \subfigure[]{\includegraphics[width=0.15\textwidth]{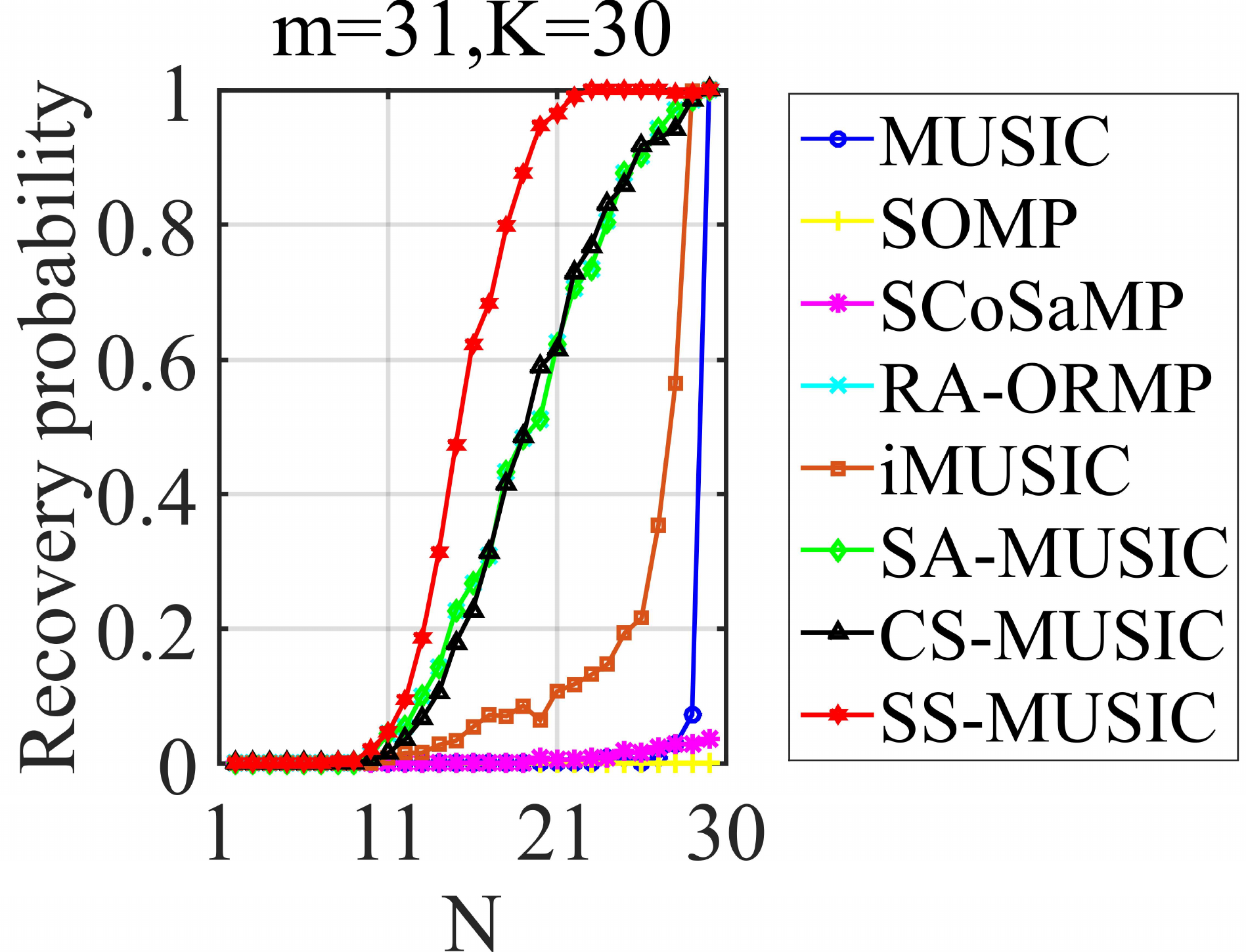}\label{RecoverProb_N}}
      \subfigure[]{\includegraphics[width=0.15\textwidth]{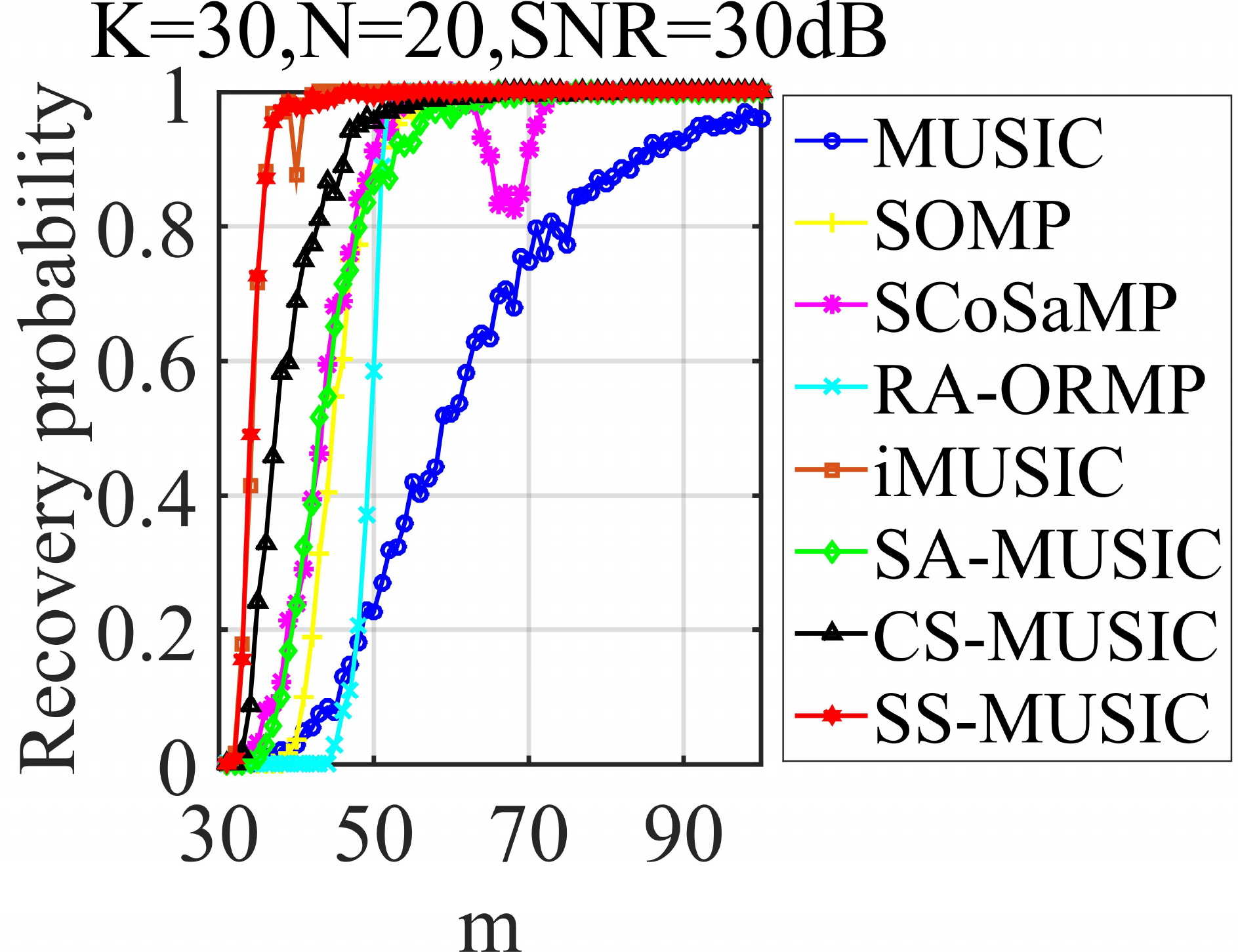}\label{RecoverProb_measurement_noise}}
   \subfigure[]{\includegraphics[width=0.15\textwidth]{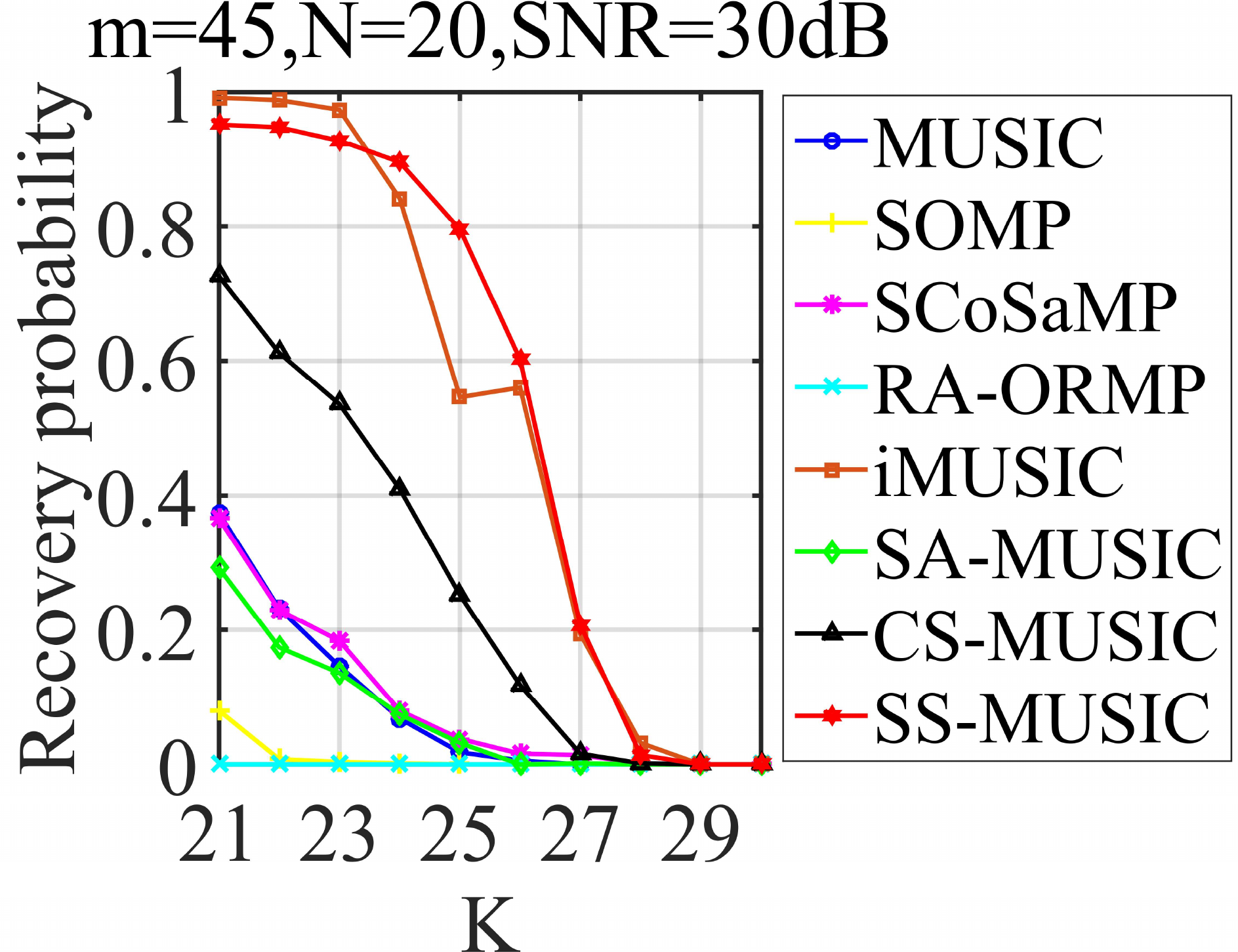}\label{RecoverProb_Sparsity_noise}}
    \subfigure[]{\includegraphics[width=0.15\textwidth]{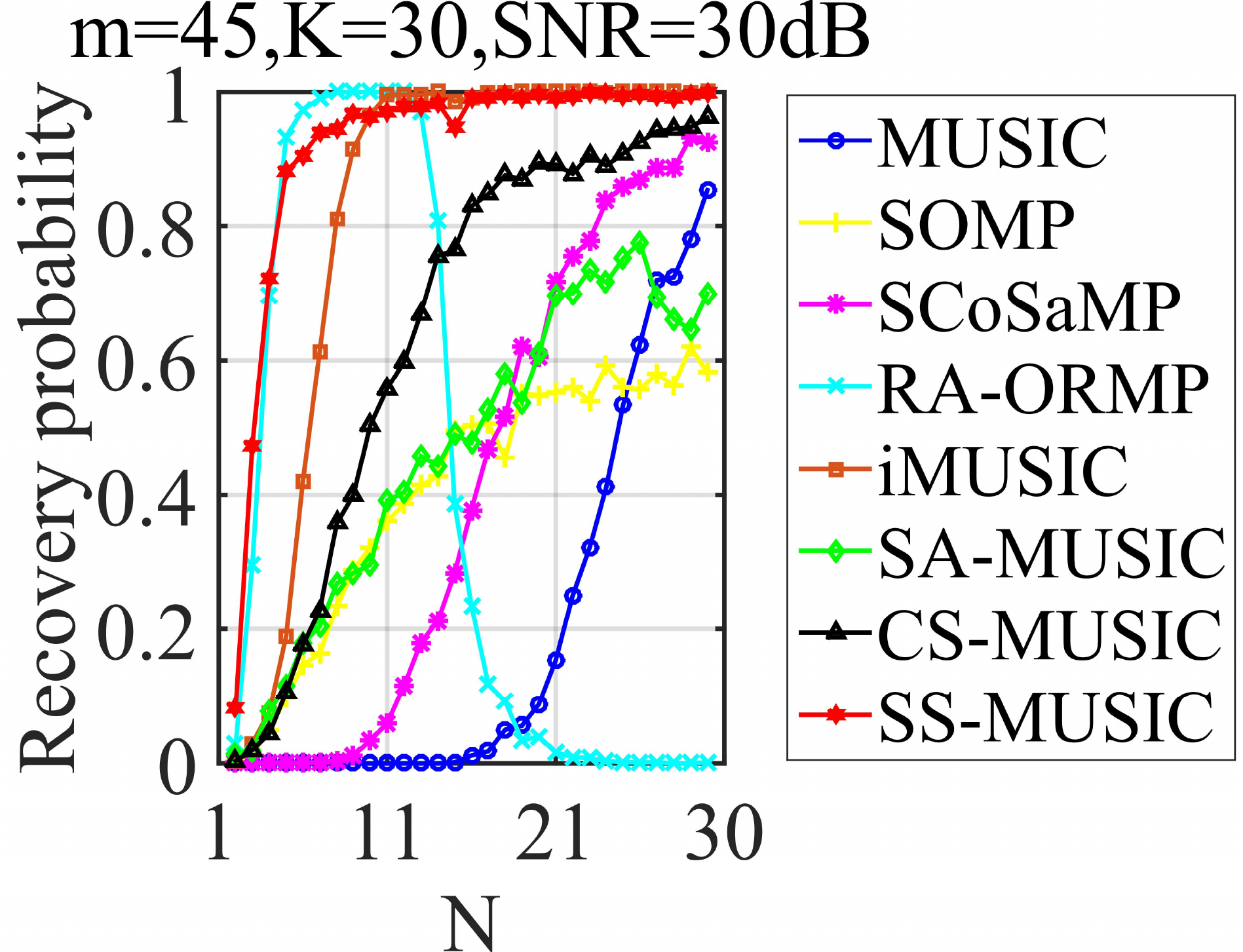}\label{RecoverProb_N_noise}}
  \caption{Comparison results of recovery probabilities in noiseless (a-c) and noisy (d-f) JSR problem with varying (a),(d). $m$; (b),(e). $K$; (c),(f). $N$.}\label{Fig:FixedSparsity}
\end{figure}

 In the next experiments, SS-MUSIC will be compared with other JSR algorithms to demonstrate its superiority in terms of required condition. For this purpose, we will vary one parameter of $m$, $K$ and $N$ to evaluate the recovery probabilities of each algorithm with the other two fixed, respectively. In the first place, the noiseless situation is considered, yielding the results shown in Figs. \ref{RecoverProb_measurement}-\ref{RecoverProb_N} and the corresponding settings are also illustrated in the figures. From the results in Fig. \ref{RecoverProb_measurement}, we can see that SS-MUSIC outperforms all competitive algorithms when we vary $m$, followed by RA-ORMP, SA-MUSIC and CS-MUSIC. It is worth noting that when $m=31=K+1$ reaches its lower bound, SS-MUSIC can still achieve over $90\%$ probability of exact recovery, which empirically verifies a milder condition of $\mathbf{A}$ for SS-MUSIC. Similar conclusions can be also derived from the performances in Figs. \ref{RecoverProb_sparsity} and \ref{RecoverProb_N} and we will not discuss for the sake of space limitation.
\par  Finally, we will evaluate their performances in noisy situation, where the measurement noise distributed from Gaussian will be considered for simplicity, namely $\mathbf{Y}_{\mathrm{noise}}=\mathbf{Y}+\mathbf{E}$ and $\mathbf{E}$ stands for the measurement noise matrix. Before starting, since SS-MUSIC is developed for noiseless MMVs, some operations should be modified, where the rank and the subspace of $\mathbf{Y}$ should be estimated from $\mathbf{Y}_{\mathrm{noise}}$ in the first stage. For simplicity and fair comparison, we adopt an efficient method proposed in SA-MUSIC to address this issue. Then we will exploit the estimated rank and basis of subspace $\mathbf{Y}$ to construct the semi-supervised classifier in SS-MUSIC. The comparison results are illustrated in Figs. \ref{RecoverProb_measurement_noise}-\ref{RecoverProb_N_noise}. We can conclude that SS-MUSIC can still preserve its remarkable recovery performance to outperform the other algorithms in the most cases. It can be also observed that iMUSIC will also achieve the better performances than other algorithms due to its refinement procedure.
\section{Conclusion}
\label{sec:Conclusion}
This paper develops a novel SS-MUSIC for rank defective JSR, which brings the strategy of ML into the optimization problem to shed a new light on this direction. We show that simultaneously exploiting the labeled MMVs and some unlabeled atoms can significantly improve the performance in terms of required iterations and conditions. In our future work, we will develop a robust framework to address the noisy JSR problem straightforwardly, in which more ML strategies will be considered and involved to avoid estimating $r$ and subspace of $\mathbf{Y}$ in the first place.

\clearpage

\bibliographystyle{IEEEtran}
\bibliography{IEEEabrv,egbib}




\end{document}